# Numerical Derivative-based Flexible Integration Algorithm for Power Electronic Systems Simulation Considering Nonlinear Components

Han Xu, *Student Member, IEEE*, Bochen Shi, *Member, IEEE*, Zhujun Yu, *Graduate Student Member, IEEE*, Jialin Zheng, *Graduate Student Member*, *IEEE*, and Zhengming Zhao, *Fellow*, *IEEE*

*Abstract*—Simulation is an efficient tool in the design and control of power electronic systems. However, quick and accurate simulation of them is still challenging, especially when the system contains a large number of switches and state variables. Conventional general-purpose integration algorithms assume nonlinearity within systems but face inefficiency in handling the piecewise characteristics of power electronic switches. While some specialized algorithms can adapt to the piecewise characteristics, most of these methods require systems to be piecewise linear. In this article, a numerical derivative-based flexible integration algorithm is proposed. This algorithm can adapt to the piecewise characteristic caused by switches and have no difficulty when nonlinear non-switching components are present in the circuit. This algorithm consists of a recursive numerical scheme that obtains high-order time derivatives of nonlinear components and a decoupling strategy that further increases computational efficiency. The proposed method is applied to solve a motor derive system and a large-scale power conversion system (PCS) to verify its accuracy and efficiency by comparing experimental waveforms and simulated results given by commercial software. Our proposed method demonstrates several-fold acceleration compared to multiple commonly used algorithms in Simulink.

## I. Introduction

SIMULATION plays a vital role in the design and verification of power electronic systems. In the simulation of power electronic systems, two types of nonlinear components are typically encountered: piecewise linear (PWL) components and smooth nonlinear components.

Switches are commonly modeled using two-value models due to their simplicity and acceptable accuracy [1]–[3]. In this way, switches are PWL components [4], [5]. There are also smooth nonlinear components such as photovoltaic arrays and motors, which do not exhibit piecewise characteristics. The coexistence of these two types of nonlinear components makes the system have both piecewise characteristic and nonlinearity and poses challenges for existing integration algorithms.

Conventional general-purpose integration algorithms assume that the state equations can be nonlinear in general and have no difficulty in solving smooth nonlinear state equations [6], [7]. However, these algorithms encounter inefficiencies when dealing with piecewise characteristics due to their limited flexibility. The algorithms can be broadly classified into two categories: the single-step variable-step and fixed-order (VSFO) algorithms, and the multi-step variable-step and variable-order (VSVO) algorithms [8], [9]. The VSFO algorithms use a fixed integration order and cannot adapt to varying lengths of continuous intervals, while the multi-step VSVO algorithms require repeated restarts after switching events happen [10].

To deal with the piecewise characteristic efficiently, specialized algorithms have been developed, resulting in a considerable speedup compared to conventional integration algorithms. However, most of them do not address the issue of nonlinearity introduced by smooth nonlinear components. Efficient matrix exponential approximations are used in [11]–[13] to achieve this speedup. Chebyshev series-based and Fourier series-based algorithms are proposed in [14], [15], respectively. Taylor series-based algorithms which can adapt to the piecewise characteristic are adopted in [16], [17]. In conclusion, these algorithms exploit the systems' linearity in some way and require systems to be PWL.

The necessity of PWL poses a constraint on the applicability of specialized integration algorithms. In the presence of smooth nonlinear components, the PWL requirement is not satisfied, and linearization techniques have been proposed to address this issue [18], [19]. However, applying these techniques in modeling multi-input multi-output components is challenging.

To overcome this limitation, decoupling methods, such as the zero-order approximation method [20] and the latency-based method [21], have been proposed to separate the system into PWL and nonlinear parts, ensuring that the PWL parts can be accelerated by specialized algorithms. Unfortunately, these decoupling methods are less accurate and stable than conventional integration algorithms.

To summarize, conventional general-purpose integration algorithms can be applied to solve systems containing smooth nonlinear components, but they are inefficient in handling the piecewise characteristics caused by switches. Specialized integration algorithms are flexible in dealing with switches but are restricted to PWL systems. This paper aims to propose an integration algorithm that is efficient in managing the piecewise characteristics caused by switches and is also applicable to solving piecewise nonlinear systems. Then this algorithm can efficiently simulate power electronics systems

This work was supported by the State Key Program of National Natural Science Foundation of China under Grant U2034201 (Corresponding author: Zhengming Zhao).

that contain nonlinear components.

To achieve this, Taylor series-based algorithms are promising solutions. These algorithms can efficiently deal with the piecewise characteristic, as they can adaptively adjust integration order and step size using high-order time derivatives (e.g., for VSVO characteristics) and have no restart issues like multi-step VSVO algorithms [16]. However, the problem of Taylor series-based algorithms is the difficulty of obtaining high-order time derivatives. Because, driving high-order derivatives' expressions for nonlinear component is impractical, and the computational complexity grows exponentially with the order [22]. To address this, automatic differentiation (AD) techniques have been adopted, as in [23]–[26]. AD is a recursive procedure that computes the derivatives of certain functions up to arbitrarily high order and is already implemented in many existing software packages [27]. While numerical derivative-based methods only require evaluating the functions in the ODEs and are easier to implement [28], [29]. However, the order of derivatives is limited in [28]. The algorithm in [29] does not provide the flexibility to adjust integration order and step size, a key advantage of Taylor series-based methods. In conclusion, current Taylor series-based algorithms are impractical when being applied to solve piecewise nonlinear system for the difficulty of deriving high-order derivatives.

In this paper, a numerical derivative-based flexible integration algorithm for power electronic systems simulation is proposed. This algorithm consists of a novel recursive numerical scheme that can easily obtain arbitrary high order time derivatives and a decoupling strategy to decouple the whole system into PWL and nonlinear parts. This algorithm can achieve efficiently simulating power electronics systems containing nonlinear components (e.g., piecewise nonlinear systems). The key advantages of this method are as follows:

1) The advantages of Taylor series-based methods are maintained. The adaptive order selection and step-size adjustment can be realized to deal with the piecewise characteristic caused by switches efficiently.
2) This recursive numerical scheme only needs to evaluate the state equations in the ODEs to derive arbitrary high-order time derivatives of smooth nonlinear components.
3) Decoupling the whole system into the PWL and nonlinear parts lowers the overall single-step computational cost, and the accuracy is ensured by exchanging numerical high-order derivatives.

It is worth noting the mentioned numerical integration algorithms are used in an event-driven framework where discrete switching event time points are located beforehand, and integration is done from current point to event time [16], [30], [31]. Another approach [32], [33] adopts a complementarity problem formulation to model power electronic systems, enabling determination of the switching states. It utilizes a fully implicit scheme for integrating and satisfying the unilateral constraints in discrete time. To maintain focus, this paper primarily aims to develop an integration algorithm that effectively handles nonlinearities, with less emphasis on the mechanism for locating the switching event time points.

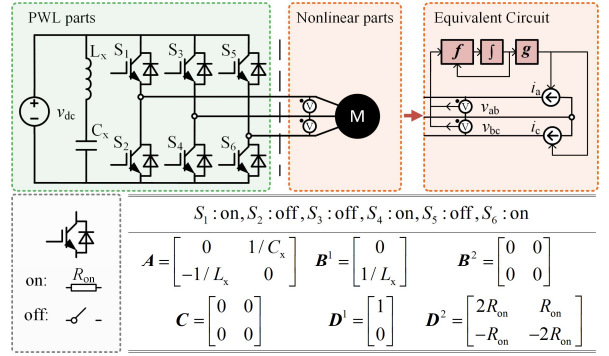

Fig. 1. An Illustrative Example of a Motor Connected to an Inverter.

The rest of this paper is organized as follows. Section II presents the modeling of power electronics systems and proposes the recursive numerical scheme for calculating high-order time derivatives and the decoupling strategy. Section III introduces the numerical derivative-based flexible algorithm and analyzes its truncation error and stability. Section IV demonstrates case studies and result analyses, and conclusions are drawn in Section V.

## II. NUMERICAL SCHEME FOR DERIVATIVES CALCULATION

In this section, the proposed recursive numerical scheme and the decoupling strategy are introduced. In addition, the numerical derivatives' errors are thoroughly analyzed.

### A. Modeling of Power Electronic Systems with Nonlinear Components

Nonlinear components can be modeled as a general state equation form as

$$\begin{aligned} \dot{\boldsymbol{x}}_{nl} &= \boldsymbol{f}(\boldsymbol{x}_{nl}, \boldsymbol{u}_{nl}) \\ \boldsymbol{y}_{nl} &= \boldsymbol{g}(\boldsymbol{x}_{nl}, \boldsymbol{u}_{nl}) \end{aligned} \quad (1)$$

where the subscript nl identifies the variables belong to nonlinear components, $\boldsymbol{x}_{nl}$ is an $n_{nl} \times 1$ vector consisting of a set of state variables, $\boldsymbol{u}_{nl}$ is a $l_{nl} \times 1$ vector representing the inputs of this component, $\boldsymbol{y}_{nl}$ is an $m_{nl} \times 1$ vector representing the outputs of this component, $\boldsymbol{f}: \boldsymbol{R}^{n_{nl}} \times \boldsymbol{R}^{l_{nl}} \mapsto \boldsymbol{R}^{n_{nl}}$ is the state function, and $\boldsymbol{g}: \boldsymbol{R}^{n_{nl}} \times \boldsymbol{R}^{l_{nl}} \mapsto \boldsymbol{R}^{m_{nl}}$ is the output function. Output vector $\boldsymbol{y}_{nl}$ and input vector $\boldsymbol{u}_{nl}$ are interface variables that connect the nonlinear components with the outside power electronic circuits.

The rest part of the system is PWL, because each switching state corresponds to a topology and a set of system matrices. Once the switching state changes, the system matrices changes, resulting in the piecewise feature of the system.

To derive the state equations of the PWL parts, each element of the input vector $\boldsymbol{u}_{nl}$ is interpreted as a voltage or current sensor. Concurrently, every element of the output vector $\boldsymbol{y}_{nl}$ is taken as a controlled voltage or current source, as depicted in Fig. 1. Then, the system matrices of the PWL part can be constructed. The detailed process for constructing the system matrices can be found in Chapter 8 of [34]. The PWL parts' equations are presented as

$$\dot{x}_1 = A_k x_1 + \begin{bmatrix} B_k^1 & B_k^2 \end{bmatrix} \begin{bmatrix} u_I \\ y_{nl} \end{bmatrix}$$
$$u_{nl} = C_k x_1 + \begin{bmatrix} D_k^1 & D_k^2 \end{bmatrix} \begin{bmatrix} u_I \\ y_{nl} \end{bmatrix} \quad (2)$$

where the subscript l identifies the variables related to the PWL parts, $x_1$ is an $n_1 \times 1$ vector consisting of a set of independent state variables, $u_I$ is a $l_1 \times 1$ vector representing the $l_1$ independent inputs, vector $u_{nl}$ represents not only the nonlinear part's input variables but also the PWL part's output variables, vector $y_{nl}$ represents not only the PWL part's input variables but also the nonlinear part's output variables, $A_k, B_k^1, B_k^2, C_k, D_k^1, D_k^2$ are system matrices that correspond to the $k$th permutation of switches. In conclusion, (1) and (2) together represent the whole system, and the whole system consists of the PWL and nonlinear parts as shown in Fig. 1.

After combining (1) and (2), there could exist an algebraic loop introduced by the output functions of the PWL and nonlinear parts. This algebraic loop can be solved analytically or numerically. In our implementation, the trust-region-dogleg algorithm [35] is adopted to solve the possible algebraic loop. Assuming the algebraic equation can be solved, the output variables $y_{nl}$ can be given by

$$y_{nl} = \hat{g}(x_{nl}, x_1, u_I) \quad (3)$$

where $\hat{g}$ represents the analytical or numerical function by solving the algebraic equation. Without loss of generality, the nonlinear parts can be modeled as

$$\dot{x}_{nl} = f(x_{nl}, u_{nl})$$
$$y_{nl} = \hat{g}(x_{nl}, x_1, u_I) \quad (4)$$

Finally, the whole system is modeled by (2) and (4).

*B. Numerical Derivatives of Nonlinear Parts*

The utilization of a Taylor series-based integration algorithm necessitates the acquisition of high-order time derivatives of state variables. However, obtaining expressions for the high-order time derivatives of nonlinear parts is generally challenging. To address this issue, a numerical scheme derived from a generalized form of state equations is introduced in this section to compute the numerical values of the time derivatives of nonlinear parts. This scheme is easy to implement since only the evaluation of functions in the ODEs is required.

Finite difference approximations techniques are needed to derive this scheme. We use $\Delta_h^{q_{max},i}$ to denote a central difference operator that can approximate $i$th-order derivatives to order $q_{max}-i$ with small value $h$. For any sufficiently differentiable function $u(t)$, it satisfies

$$h^i \Delta_h^{q_{max},i} u(t_0 + t) = h^i u^{(i)}(t_0) + O(h^{q_{max}}). \quad (5)$$

Without loss of generality, we start from a scalar autonomous ordinary differential equation as

$$x^{(1)} = f(x) \quad (6)$$

where $x$ is a state variable and $\dot{x}$ denotes the first-order time derivative of $x$. At $t = t_0$, it is assumed we have $x = x_0$.

The numerical result of the first order time derivative of state variable $x$ is given as

$$\tilde{x}_0^{(1)} = x_0^{(1)} = f(x_0). \quad (7)$$

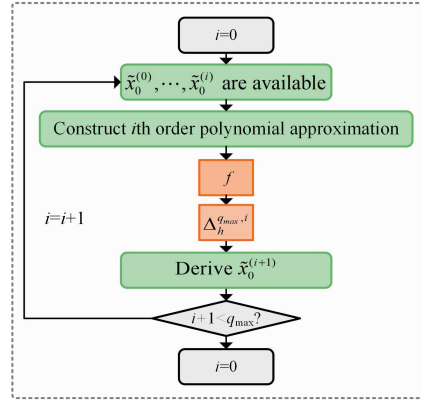

Fig. 2. Illustration of the proposed recursive numerical scheme.

The derivation of the numerical scheme starts from a novel observation of (7). It is natural to interpret that (7) is obtained by directly using (6). However, it can also be interpreted as applying the operator $\Delta_h^{q_{max},0}$ to $x_0$, the zero-order approximation to $x(t)$ at point $t_0$

$$\tilde{x}_0^{(1)} = \Delta_h^{q_{max},0} f(x_0). \quad (8)$$

After getting the first order derivative, the first-order approximation to $x(t)$ can be constructed as

$$\tilde{P}_{x,1}(t) = \tilde{x}_0 + \tilde{x}_0^{(1)}(t - t_0) \quad (9)$$

where $\tilde{x}_0 = x_0$.

The above process can be done recursively to obtain higher-order time derivatives. The recursive formula for calculating numerical time derivatives is given as

$$\begin{cases} h^i \tilde{x}_0^{(i+1)} = h^i \Delta_h^{q_{max},i} f(\tilde{P}_{x,i}(t+t_0)) \\ \tilde{P}_{x,i}(t) = \sum_{k=0}^{i} \frac{\tilde{x}_0^{(k)}}{k!}(t-t_0)^k \end{cases}, i = 0, \cdots, q_{max}-1 \quad (10)$$

where $\tilde{P}_{x,i}(t)$ is the $i$th order polynomial approximation to $x(t)$ at point $t_0$ and the coefficients of this polynomial is determined by the numerical result $\tilde{x}_0^{(k)}$.

Even though this numerical scheme is derived from scalar functions, it can be applied to calculate vector functions' derivatives by simply substituting the scalar functions with vector functions and the numerical results of time derivatives of the state function $f(x_{nl}, u_{nl})$ and the output function $\hat{g}(x_{nl}, x_1, u_I)$ are given as

$$h^i \tilde{x}_{nl}^{(i+1)} = h^i \varDelta_h^{q_{max},i} f(\tilde{P}_{x_{nl},i}(t+t_0), \tilde{P}_{u_{nl},i}(t+t_0)) \quad (11)$$

$$h^i \tilde{y}_{nl}^{(i)} = h^i \varDelta_h^{q_{max},i} \hat{g}(\tilde{P}_{x_{nl},i}(t+t_0), \tilde{P}_{x_1,i}(t+t_0), \tilde{P}_{u_I,i}(t+t_0)) \quad (12)$$

$$\tilde{P}_{x_{nl},i}(t) = \sum_{k=0}^{i} \frac{\tilde{x}_{nl}^{(k)}}{k!}(t-t_0)^k, \tilde{P}_{x_1,i}(t) = \sum_{k=0}^{i} \frac{\tilde{x}_1^{(k)}}{k!}(t-t_0)^k,$$
$$\tilde{P}_{u_{nl},i}(t) = \sum_{k=0}^{i} \frac{\tilde{u}_{nl}^{(k)}}{k!}(t-t_0)^k, \tilde{P}_{u_I,i}(t) = \sum_{k=0}^{i} \frac{\tilde{u}_I^{(k)}}{k!}(t-t_0)^k \quad (13)$$

where $\varDelta_h^{q_{max},i}$ is the generalization of the differentiation operator $\Delta_h^{q_{max},i}$ to vector functions.

The errors of the numerical results given by (11) and (12) obey

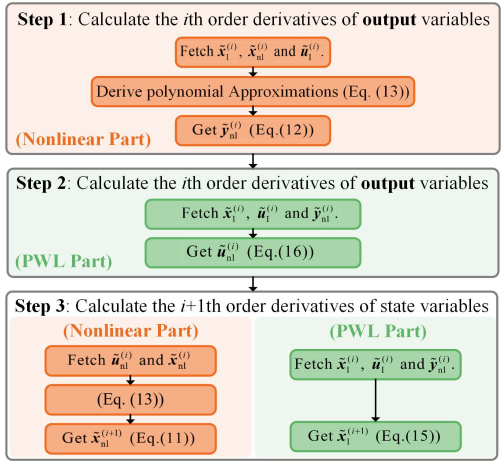

Fig. 3. Illustration of the proposed decoupling strategy.

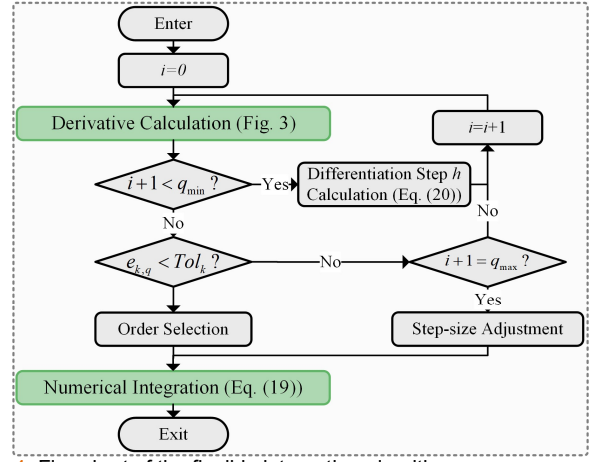

Fig. 4. Flowchart of the flexible integration algorithm.

$$h^i \tilde{x}_{nl}^{(i+1)} - h^i x_{nl}^{(i+1)} = O(h^{q_{max}})$$
$$h^i \tilde{y}_{nl}^{(i)} - h^i y_{nl}^{(i)} = O(h^{q_{max}}), 0 \leq i \leq q_{max} - 1 \quad (14)$$

when the errors of $h^k \tilde{x}_{nl}^{(k)}$, $h^k \tilde{u}_{nl}^{(k)}$, $h^k \tilde{x}_1^{(k)}$ and $h^k \tilde{u}_1^{(k)}$ are at least $q_{max}$th order for $k = 0, \cdots, i$. This relation is proved in the appendix.

### C. Numerical Derivatives of PWL Parts and Decoupling Strategy

In this section, we demonstrate a simple recursive formula to obtain high-order time derivatives of the PWL parts. Additionally, we propose a decoupling strategy that allows the use of the simple recursive formula on the PWL parts while the nonlinear parts are calculated using equations (11) and (12).

Differentiating both sides of the PWL parts' equations in (2), and substituting the true values with numerical values, the numerical scheme for calculating the PWL parts' time derivatives is given as

$$\tilde{x}_1^{(i+1)} = A_k \tilde{x}_1^{(i)} + \begin{bmatrix} B_k^1 & B_k^2 \end{bmatrix} \begin{bmatrix} \tilde{u}_1^{(i)} \\ \tilde{y}_{nl}^{(i)} \end{bmatrix} \quad (15)$$

$$\tilde{u}_{nl}^{(i)} = C_k \tilde{x}_1^{(i)} + \begin{bmatrix} D_k^1 & D_k^2 \end{bmatrix} \begin{bmatrix} \tilde{u}_1^{(i)} \\ \tilde{y}_{nl}^{(i)} \end{bmatrix}. \quad (16)$$

Combining (11)-(13), (15) and (16), a decoupling strategy for calculating derivatives of the whole system is proposed and illustrated in Fig. 3. The flowchart in Fig. 3 shows how to derive the $i+1$th order time derivatives when the lower-order derivatives are ready. Initially, state variables $\tilde{x}_{nl}^{(k)}$, $\tilde{x}_1^{(k)}$ for $k = 0, \cdots, i$ and interface variables $\tilde{u}_{nl}^{(k)}$, $\tilde{y}_{nl}^{(k)}$ for $k = 0, \cdots, i-1$ are available. The time derivatives of $\tilde{u}_1$ are always available, since this input vector is composed of explicit independent sources such as sinusoidal sources with harmonics and dc sources, whose time derivatives can be easily evaluated.

In step 1, the $i$th-order time derivatives of output variables $\tilde{y}_{nl}^{(i)}$ of the nonlinear parts are derived through (12). In step2, after $\tilde{y}_{nl}^{(i)}$ is ready, the $i$th-order time derivatives of the output variables $\tilde{u}_{nl}^{(i)}$ of the PWL parts can be derived through (16). In step 3, interface variables are available. Therefore, the PWL and nonlinear parts can calculate their own time derivatives of state variables respectively through (11) and (15).

### D. Numerical Derivatives Error Analysis

According to the proposed numerical scheme, the nonlinear parts' time derivatives have errors, and they are also used for the calculation of the PWL parts' time derivatives. In this part, we answer if the errors will propagate.

For the PWL parts, the time derivatives' numerical errors are shown as

$$h^i \tilde{x}_1^{(i+1)} - h^i x_1^{(i+1)} = A_k (h^i \tilde{x}_1^{(i)} - h^i x_1^{(i)}) + B_k^2 (h^i \tilde{y}_{nl}^{(i)} - h^i y_{nl}^{(i)})$$
$$h^i \tilde{u}_{nl}^{(i)} - h^i u_{nl}^{(i)} = C_k (h^i \tilde{x}_1^{(i)} - h^i x_1^{(i)}) + D_k^2 (h^i \tilde{y}_{nl}^{(i)} - h^i y_{nl}^{(i)}) \quad (17)$$

The zero-order values of all state variables and output variables have no errors, and by induction, we can combine (14) and (17) to prove

$$h^i \tilde{x}_1^{(i+1)} - h^i x_1^{(i+1)} = O(h^{q_{max}})$$
$$h^i \tilde{u}_{nl}^{(i)} - h^i u_{nl}^{(i)} = O(h^{q_{max}}), 0 \leq i \leq q_{max} - 1 \quad (18)$$

## III. FLEXIBLE INTEGRATION ALGORITHM WITH NUMERICAL DERIVATIVES

### A. The Flowchart of the Flexible Integration Algorithm

In this section, a numerical derivative-based flexible integration algorithm is proposed. This algorithm combines the recursive numerical scheme and the decoupling strategy introduced in Section II. The complete flowchart of the integration algorithm proposed in this paper is presented in Fig. 4. This algorithm uses the truncated Taylor series expansion of state variables to approximate their solutions at time point $t_{k+1}$

$$\tilde{x}_{s,k+1} = x_{s,k} + \sum_{i=1}^{q} \frac{\tilde{x}_{s,k}^{(i)}}{i!} h_{step,k}^i \quad (19)$$

where $\tilde{x}_{s,k}^{(i)}$ denotes the $i$th order numerical time derivatives, the subscript $s$ represents nl or l, $q$ denotes the chosen integration order and $h_{step,k}$ denotes the step-size of the $k$th step.

This algorithm departs from the conventional Taylor series-based methods by utilizing numerical derivatives instead of analytical derivatives. To obtain these numerical derivatives, the small value $h$ must be determined first for the numerical proposed numerical scheme. As shown in Fig. 4, there is a differentiation step calculation stage. We employ a practical approach to determine this value. Since the first order integration method is impractical due to its step-size limitation, we set the minimum integration order $q_{min}$ to be the second order. Furthermore, we employ the step size obtained for the first order integration method as the small value $h$ to conduct the numerical differentiation

$$h_k = \frac{Tol_k}{\max(\|x_{nl}^{(1)}\|_\infty, \|x_1^{(1)}\|_\infty)} \quad (20)$$

where $\|\cdot\|_\infty$ denotes infinite norm, $Tol_k$ denotes the error tolerance of the $k$th step, and $h_k$ denotes the small value for numerical differentiation of the $k$th step.

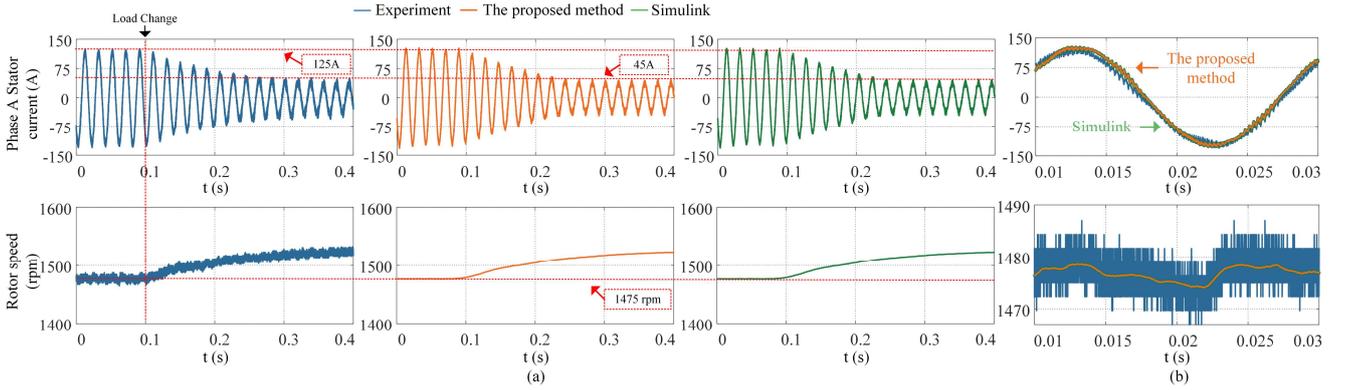

Fig. 5. Comparison between experimental and simulated waveforms. (a) Motor's stator current of phase A and rotor speed. (b) Zoomed-in view

Once the small value for numerical differentiation is chosen, the higher-order time derivatives of the nonlinear and PWL parts can be obtained through the decoupling strategy shown in Fig. 3. The integration order and step-size are chosen based on the truncation error. The truncation error of the $q$th order is given by

$$e_{k,q} = (\frac{\|x_k^{(q)}\|_\infty}{q!})^{\frac{q+1}{q}} h_{step,k}^{q+1}. \quad (21)$$

The proposed algorithm is efficient due to its flexibility in adaptively adjusting the step size and order, as elaborated in the case study section. Furthermore, the adjustments to the step size and order do not require any additional computational cost. For step-size adjustment, derivatives can be reused for numerical integration, eliminating the need for recalculating them. For order adjustment, derivatives can be used for both numerical integration and truncation error estimation. In contrast, Runge-Kutta methods require reintegrating when reducing the step size to meet accuracy requirements.

The efficiency of this integration method also relies on the low single-step computational cost. The comparisons between the single-step computational cost of the proposed method and that of the conventional integration algorithms are shown in Table I. This comparison also reveals the decoupling strategy lowers the overall computational cost of calculating numerical derivatives. Because calculating the PWL parts' derivatives is less computationally intensive.

### B. Truncation Error and Numerical Stability Analysis

The truncation error of the flexible algorithm consists of two parts: the error caused by Taylor series truncation and the error introduced by numerical derivatives. Combining (14), (18), and (19), the truncation error of the $q$th order integration is given as

TABLE I
COMPUTATIONAL COST COMPARISONS

| Algorithms | Computational cost per step | |
|---|---|---|
| | PWL parts (Number of multiplications) | Nonlinear parts (Number of function evaluations) |
| Dormand-Prince algorithm | $7n_1(n_1+l_1+m_{nl})+O(n_1)$ | 7 |
| Bogacki–Shampine method | $4n_1(n_1+l_1+m_{nl})+O(n_1)$ | 4 |
| The proposed method (order = 5) | $5n_1(n_1+l_1+m_{nl})+O(n_1)$ | 17 |
| The proposed method (order = 3) | $3n_1(n_1+l_1+m_{nl})+O(n_1)$ | 9 |

*Note:* The Dormand-Prince algorithm achieves six function evaluations per step by exploiting the FSAL (First Same as Last) property. However, discrete events limit the continuous integration range, resulting in an average of seven function evaluations. Similarly, the Bogacki-Shampine (BS) algorithm, the average number of function evaluations is around four.

$$\begin{aligned} E_k &= x_{s,k+1} - (x_{s,k} + \sum_{i=1}^{q} \frac{\tilde{x}_{s,k}^{(i)}}{i!} h_{step,k}^i) \\ &= x_{s,k+1} - (x_{s,k} + \sum_{i=1}^{q} \frac{x_{s,k}^{(i)}}{i!} h_{step,k}^i) + (\sum_{i=1}^{q} \frac{x_{s,k}^{(i)} - \tilde{x}_{s,k}^{(i)}}{i!} h_{step,k}^i) \\ &= \frac{x_{s,k}^{(q+1)}}{(q+1)!} h_{step,k}^{q+1} + O(h_{step,k}^{q+2}) + (\sum_{i=1}^{q} \frac{O(h_k^{q_{max}+1})}{i!} h_{step,k}^i / h_k^i) \\ &= O(h_{step,k}^{q+1}) + O(h_{step,k}^{q_{max}+1}). \end{aligned} \quad (22)$$

The lower bound $q_{min}$ ensures the integration step $h_{step,k}$ is greater than the differentiation step $h_k$. The upper bound ensures the order of the integration method is always less than or equal to that of the error introduced by numerical derivatives. Once $q$ is within $q_{min}$ and $q_{max}$, the errors introduced by numerical derivatives are always compatible with the order of the integration algorithm and the integration error can be well controlled.

Regarding numerical stability, the proposed method is based on Taylor expansion, and the stability function of the proposed $q$th-order method is the same as that of the $q$th-order Runge-Kutta method, which is also based on Taylor expansion.

### C. Discussions

The novelty of the numerical derivative-based flexible integration method proposed in this paper stems from the proposed recursive numerical scheme and the decoupling strategy. Compared to the existing methods [28], [29], our proposed recursive numerical scheme can compute numerical derivatives to arbitrarily high order, and it only requires the evaluation of the functions in the ODEs. Furthermore, the numerical error introduced by our method for the derivatives has been proven in the paper to be compatible with the integration error. In terms of the decoupling strategy, the interface variables used in our proposed decoupling strategy take advantage of high-order numerical derivatives. In contrast to latency-based decoupling methods [21] and those requiring linear systems [16], our method not only ensures high precision but is also applicable to nonlinear systems.

## IV. CASE STUDY AND RESULT ANALYSIS

In this section, two cases are simulated using the proposed flexible integration algorithm to verify its accuracy and efficiency. The first case involves a 55kW/380V induction motor drive system, while the second case involves a 10kV PCS converter that is connected to renewable sources, a motor driving load, and the main grid.

### A. Case 1: Induction Motor Drive System

The proposed method is applied to solve a 55kW/380V induction motor drive system. The circuit of this system is shown in Fig. 1. The motor is driven by a three-phase two-level inverter, and the DC source's voltage is 480V. The field-orientated control (FOC) strategy is adopted for controlling the induction motor.

In this experimental study, the motor is operated under closed-loop control and starts without any load. It gradually attains the reference rotor speed of 1475 r/min and then supports a rated load of 350 Nm. Subsequently, the load is reduced to 170 Nm. The dynamic process of load-down is shown through simulated and experimental waveforms, as presented in Fig. 5. The simulation results are compared with experimental results and results given by commercial software to verify the accuracy of the proposed flexible algorithm.

As shown in Fig. 5, the waveforms of the current and rotor speed of the simulation and experiment are in good agreement. Although there are speed ripples in the experimental speed waveform, and the simulated speed waveforms are more ideal and smoother, the average rotor speed matches. Drawing upon the zoomed-in view, it is evident that the noticeable fluctuations in rotor speed are partly attributed to the sampling discretization of the actual speed value.

### B. Case 2: PCS

Fig. 6 illustrates the overall structure of the simulated PCS. The DC side connects with PV arrays, batteries, DC loads, and a motor load. The AC side connects with AC loads and the main grid. PV arrays [36], batteries [37] and motor loads are smooth nonlinear components. These nonlinear components' equivalent circuits are shown in Fig. 6 and the detailed equations can be found in the Appendix. The PWL part of this PCS system contains 138 switches and 96 states variables. The nonlinear parts include 15 batteries, 15 PV arrays and 1 motor load.

The simulation is conducted for a duration of 0.4 seconds. At 0.1s, PV arrays are connected to the DC bus, followed by the connection of DC loads to the DC bus. At 0.25 s, the system becomes islanded.

*1) Comparisons with Commercial Software*

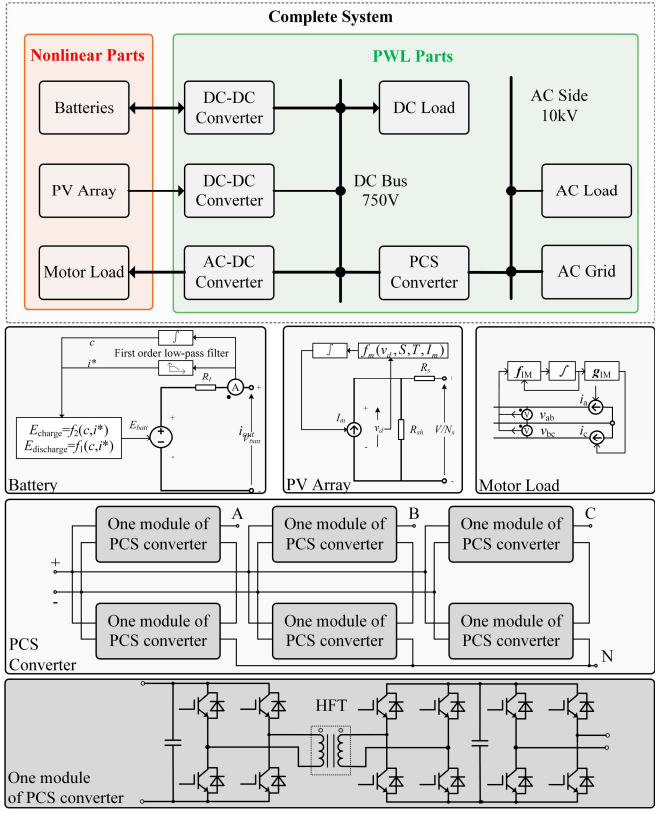

Fig. 6. Structure of the simulated PCS.

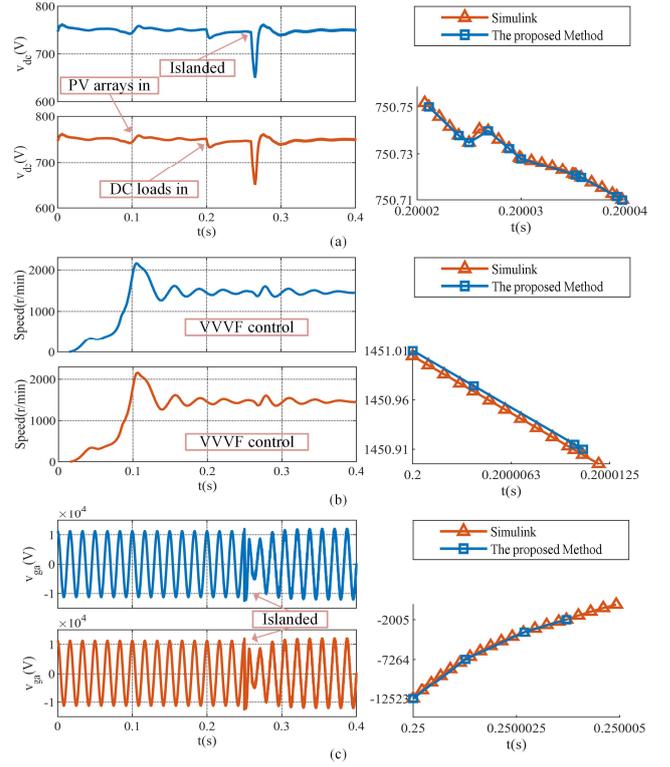

Fig. 7. Comparisons between waveforms given by the proposed method and given by Simulink. (a) DC bus voltage. (b) Rotation speed of the motor load. (c) PCS converter phase current vga.

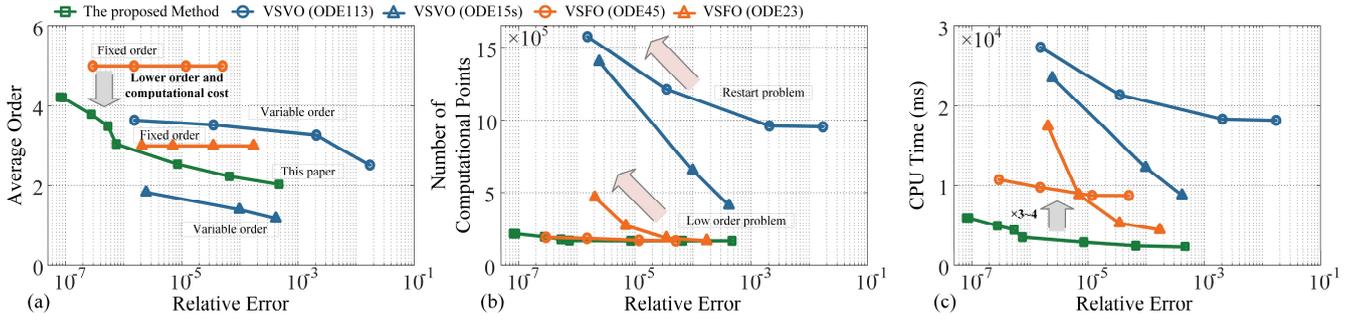

Fig. 8. Comparisons between conventional numerical integration methods and the proposed method.

TABLE II
PROGRAM PERFORMANCE PROFILE COMPARISON

| Methods | Total Time (s) | Function Evaluation (ms) | Interface Variable Calculation (ms) | Other (ms) |
|---|---|---|---|---|
| The proposed method | 3.319 | 2906 | 284 | 129 |
| ODE45 | 9.971 | 8193 | 1546 | 232 |
| Commercial Software | 3480 | - | - | - |

*Note:* The relative errors of different methods are set to be the same as 1e-4.

The reference waveforms are given by the commercial software Simulink. The relative error is set to be 1e-9 to ensure accuracy. Compared with the reference waveforms, it can be observed that the simulated results are in agreement, including the ripple of the DC bus voltage, the motor's rotation speed, and the three-phase voltage and current waveforms of the PCS converter shown in Fig. 7.

*2) Simulation Efficiency Verification*

To validate the efficiency of the proposed approach, a C++ code was implemented for several conventional numerical integration methods and the proposed method. These codes were identical except for the integration algorithm used. The simulations were executed on a personal computer equipped with a 3.1 GHz Intel Core TM i5-10500 CPU to ensure fair comparisons

$$Error_{rel} = \sum_{i=1}^{n} \frac{|y_{\text{sim},i} - y_{\text{ref},i}|}{\max(y_{\text{ref},i}, absTol)} / n \quad (23)$$

where $y_{\text{sim},i}$ is the value of an output variable at the $i$th time step, $y_{\text{ref},i}$ is the interpolation value of the reference waveform at the $i$th time step, $absTol$ is the absolute tolerant error, and $n$ is the number of all points used for comparison. Fig. 8 shows the variation of the number of time steps, the average integration order, and the integration time of different algorithms as the relative error varies. Based on the truncation error equation (22), increasing the integration order and decreasing the step-size can reduce the truncation error. This correlation can be used to analyze the results shown in Fig. 8.

Firstly, a comparison is made between the proposed method and the single-step VSFO methods, specifically ode23 and ode45. As shown in Fig. 8(a), the average order of ode23 and ode45 remains constant with respect to the relative error, while the proposed algorithm can adjust the step size and integration order based on error requirements after each discrete event occurrence. This flexibility makes the proposed method more efficient than the fixed-order algorithms under varying relative error tolerances. Specifically, when the relative error tolerance is large, the proposed method can adaptively use low-order to reduce the single-step computational cost. Conversely, when the relative error tolerance is small, the proposed method can adaptively increase the order to maintain almost the same computational points.

Secondly, the proposed method is also compared to multi-step VSVO methods (ode113 and ode15s). Fig. 8(a) demonstrates that multi-step methods increase the average integration order as the relative error decreases, but their requirement of restarting after each discrete event and enough historical values for increasing order results in significant increase of computational points. Therefore, the multi-step VSVO method may not be efficient for simulating power electronic systems in this case due to the restart problem.

In conclusion, our proposed method consistently outperforms in terms of efficiency under various relative error settings, and it shows 3 to 10-fold acceleration compared to conventional numerical integration algorithms.

### E. Comprehensive Evaluation

To comprehensively show the efficiency of our proposed method, the CPU time of different program parts is shown in Table II. As can be observed, the time consumed by function evaluations is reduced by our method. Moreover, our method does not cause large computational burdens in other parts of the program. The efficiency of the proposed method can be attributed to two reasons.

1) Flexible variable order and step size. The VSVO method can be adapted to different continuous interval lengths and relative error requirements so that the most efficient order is always chosen for each step. As a result, the number of calculated points and the average integration order (e.g., the average number of function evaluations) are reduced due to the flexibility of our method.

2) Lower single-step computational cost. The proposed method partitions the whole system into the nonlinear and the PWL part to keep the PWL parts' computational cost low. Therefore, the overall single-step computational cost of the proposed method is still lower than that of ode45 and ode23 under the same numerical accuracy order.

The limitation of the proposed method is that it is more suitable for solving non-stiff and slightly stiff systems. The proposed method can easily increase integration order when dealing with slightly stiff systems. However, the proposed method becomes inefficient when systems are highly stiff because the stability requirement limits the step size.

## V. CONCLUSION

A numerical derivative-based flexible integration algorithm is proposed to accelerating the simulation of power electronic systems containing nonlinear components. This algorithm comprises a recursive numerical scheme and a decoupling strategy. The numerical derivatives of nonlinear components are obtained through the proposed numerical scheme. The system is decoupled into nonlinear and piecewise linear (PWL) parts using the decoupling strategy, enabling their derivatives to be computed using different equations. As a result, this algorithm is flexible in adaptively adjusting step-size and order to efficiently handle the piecewise characteristics, without requiring the system to be PWL. The proposed method is applied to solve a motor drive system and a large-scale PCS to verify its accuracy and efficiency. The experimental results of the motor drive system are presented and in good agreement with the simulated results. For simulating PCS, the proposed method shows several-fold acceleration compared to multiple commonly used general-purpose algorithms in Simulink.

## APPENDIX

### A. The proof of the truncation errors

In this section, the relation given by (14) is proved.

Assume vector functions $f : R^m \mapsto R^n$ and $x : R \mapsto R^m$ are sufficiently differentiable, $y_0 = f(x_0)$, $x(t_0) = x_0$ and we have

$$h^k \tilde{x}_0^{(k)} = h^k x_0^{(k)} + O(h^{q_{max}}), \ k = 0, \cdots, i \quad (24)$$

where $h$ is some small value, $x_0^{(k)}$ are the $k$th-order derivatives with respect to $t$ at $t_0$, and $\tilde{x}_0^{(k)}$ are the corresponding approximations.

We define

$$P_{x,i}(t) = \sum_{k=0}^{i} \frac{x_0^{(k)}}{k!}(t - t_0)^k, \quad (25)$$

$$\tilde{P}_{x,i}(t) = \sum_{k=0}^{i} \frac{\tilde{x}_0^{(k)}}{k!}(t - t_0)^k. \quad (26)$$

The $i$th-order time derivatives $y_0^{(i)}$ are approximated as

$$h^i \tilde{y}_0^{(i)} = h^i \Delta_h^{q_{max},i} f(\tilde{P}_{x,i}(t + t_0)). \quad (27)$$

Then the error can be divided into three parts as

$$h^i \tilde{y}_0^{(i)} - h^i y_0^{(i)} =$$
$$(h^i \Delta_h^{q_{max},i} f(\tilde{P}_{x,i}(t + t_0)) - h^i \Delta_h^{q_{max},i} f(P_{x,i}(t + t_0))) +$$
$$(h^i \Delta_h^{q_{max},i} f(P_{x,i}(t + t_0)) - h^i f^{(i)}(P_{x,i}(t + t_0))) + \quad (28)$$
$$(h^i f^{(i)}(P_{x,i}(t + t_0)) - h^i f^{(i)}(x_0))$$

Applying Taylor expansion, the first part can be further expressed as

$$h^i \Delta_h^{q_{max},i} f(\tilde{P}_{x,i}(t + t_0)) - h^i \Delta_h^{q_{max},i} f(P_{x,i}(t + t_0))$$
$$= h^i \Delta_h^{q_{max},i} Jf(x_0)(\tilde{P}_{x,i}(t + t_0) - P_{x,i}(t + t_0)) + \quad (29)$$
$$O((\tilde{P}_{x,i}(t + t_0) - P_{x,i}(t + t_0))^2)$$

Based on (5), the second part's error is $q_{max}$th order. The third part equals zero for

$$P_{x,i}^{(k)}(t_0) = x_0^{(k)}, k = 0, \cdots, i. \quad (30)$$

From (28)-(30), the errors of the numerical results are rewritten as

$$h^i \tilde{y}_0^{(i)} - h^i y_0^{(i)} = h^i \Delta_h^{q_{max},i} Jf(x_0)(\sum_{k=0}^{i} \frac{\tilde{x}_0^{(k)} - x_0^{(k)}}{k!}t^k) +$$
$$O((\sum_{k=0}^{i} \frac{\tilde{x}_0^{(k)} - x_0^{(k)}}{k!}h^k)^2) + O(h^{q_{max}}) \quad (31)$$

Based on the assumption (24), we can prove

$$h^i \tilde{y}_0^{(i)} - h^i y_0^{(i)} = O(h^{q_{max}}). \quad (32)$$

### B. Equations of the nonlinear components

#### 1) PV Array

The state equations of the PV arrays are non-linear and are illustrated in (33), and its equivalent circuit is shown in Fig. 6. $N_s$ is the number of cells of the solar module. $S_0$ is the light intensity under the standard test conditions. $T_{ref}$ is the temperature under the standard test conditions. $R_s$ and $R_{sh}$ are series and shunt resistance of each solar cell. $I_{s0}$ is the diode saturation current of each solar cell at the reference temperature $T_{ref}$. $E_g$ is the band energy of each solar cell. $A$ is the ideality factor of each solar cell, also called emission coefficient. $C_t$ is the temperature coefficient. $K_s$ defines how the light intensity affects the solar cell temperature. The variable $I_m$ is the equivalent controlled current source shown in the equivalent circuit in Fig. 6, and $v_d$ is the voltage of the equivalent controlled current source. The variables $S$ and $T$ are input variables denoting the light intensity and the temperature respectively.

$$\frac{dI_m}{dt} = f_m(v_d, S, T, I_m)$$

$$\begin{cases} f_m(v_d, S, T, I_m) = (I_{sc0}(\frac{S}{S_0}) + C_t(T - T_{ref}) - \\ I_{s0}(\frac{T}{T_{ref}})^3 (e^{(\frac{qE_g}{Ak})(\frac{1}{T_{ref}} - \frac{1}{T})})(e^{(\frac{qv_d}{AkT})} - 1) - I_m)/t_d \\ v_d = I_m \frac{R_s R_{sh}}{R_s + R_{sh}} + V \frac{R_{sh}}{N_s(R_s + R_{sh})} \end{cases} \quad (33)$$

#### 2) Battery

The output equations of the batteries are non-linear and are illustrated in (34), and its equivalent circuit is shown in Fig. 6. $E_{Batt}$ is the controlled nonlinear voltage, as shown in Fig. 6. $E_0$ is the constant voltage. $K$ is the polarization constant also called polarization resistance. $i^*$ is the low-frequency current dynamics. $i_{out}$ is the battery current. $c$ is the extracted capacity. $Q$ is the maximum battery current. $A$ is the exponential voltage. $B$ is the exponential capacity.

$$\frac{dc}{dt} = i_{out}, \frac{di^*}{dt} = (i^* - i_{out})/T$$

$$E_{batt} = \begin{cases} f_2(c, i^*) & i^* < 0, \text{charge} \\ f_1(c, i^*) & i^* > 0, \text{discharge} \end{cases}$$

$$f_1(c, i^*) = E_0 - K\frac{Q}{Q-c}i^* - K\frac{Q}{Q-c}c + Ae^{-Bc} \quad (34)$$

$$f_2(c, i^*) = E_0 - K\frac{Q}{0.1Q+c}i^* - K\frac{Q}{Q-c}c + Ae^{-Bc}$$

## 2) Induction motor

The state equations of the three-phase squirrel-cage induction motor are nonlinear and are illustrated in (35). The variables $i_{sd}$, $i_{sq}$ and $\psi_{rd}$, $\psi_{rq}$ are stator currents and rotor flux respectively in $d$-$q$-$o$ axis. The variables $u_{sd}$, $u_{sq}$ are motor's input voltages in $d$-$q$-$o$ axis. The variable $\omega_2$ is the electrical angular velocity of rotor. $r_r$, $r_s$ are rotor resistance and stator resistance. $L_m$, $L_r$, $L_s$ are magnetizing inductance, self-inductance of rotor and self-inductance of stator, respectively. $p_0$ is the number of pole pairs of the motor. $T_L$ represents the load torque. $T_r$ is the electro-magnetic time constant of rotor. $\sigma$ is the magnetic flux leakage coefficient.

$$\begin{cases} \dfrac{di_{sd}}{dt} = \dfrac{L_m}{\sigma L_s L_r T_r}\cdot\Psi_{rd} + \dfrac{L_m}{\sigma L_s L_r}\cdot\omega_2\Psi_{rq} - \dfrac{r_s L_r^2 + r_r L_s^2}{\sigma L_s L_r^2}\cdot i_{sd} + \dfrac{u_{sd}}{\sigma L_s} \\ \dfrac{di_{sq}}{dt} = \dfrac{L_m}{\sigma L_s L_r T_r}\cdot\Psi_{rq} - \dfrac{L_m}{\sigma L_s L_r}\cdot\omega_2\Psi_{rd} - \dfrac{r_s L_r^2 + r_r L_s^2}{\sigma L_s L_r^2}\cdot i_{sq} + \dfrac{u_{sq}}{\sigma L_s} \\ \dfrac{d\Psi_{rd}}{dt} = -\dfrac{1}{T_r}\cdot\Psi_{rd} - \omega_2\Psi_{rq} + \dfrac{L_m}{T_r}\cdot i_{sd} \\ \dfrac{d\Psi_{rq}}{dt} = -\dfrac{1}{T_r}\cdot\Psi_{rq} + \omega_2\Psi_{rd} + \dfrac{L_m}{T_r}\cdot i_{sq} \\ \dfrac{d\omega_2}{dt} = \dfrac{p_0^2 L_m}{J L_r}\cdot(i_{sq}\Psi_{rd} - i_{sd}\Psi_{rq}) - \dfrac{p_0}{J}\cdot T_L \end{cases}$$

$$T_r = \dfrac{L_r}{r_r}, \quad \sigma = \dfrac{L_s L_r - L_m^2}{L_s L_r}$$

(35)

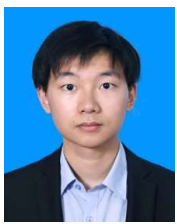
**Han Xu** (Student Member, IEEE) received the B.S. degree in electrical engineering in 2021 from Tsinghua University, Beijing, China. Since 2021, he has been working toward the master degree in electrical engineering at the Department of Electrical Engineering, Tsinghua University, Beijing, China. His research interests include simulation of power electronic systems, and parallel computing.

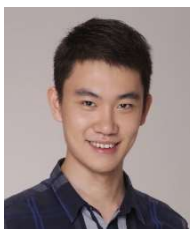
**Bochen Shi** (Member, IEEE) received the B.S. degree and the Ph.D. degree in electrical engineering from the Department of Electrical Engineering, Tsinghua University, Beijing, China, in 2017 and 2022 repectively. He is now a postdoctoral research fellow at the Department of Electrical Engineering, Tsinghua University. His research interests include modeling and simulation of power electronics systems and switching transients of power converters.

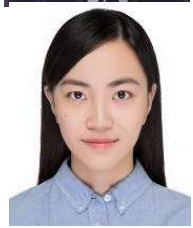
**Zhujun Yu** (Graduate Student Member, IEEE) received her B.S. degree in electrical engineering from the Department of Electrical Engineering, Tsinghua University, Beijing, China, in 2018, where she is currently working towards her Ph.D. degree in electrical engineering.
At Tsinghua University, she works on the discrete state event-driven simulation of power electronics systems and the development of DSIM. Her research interests include computer aided analysis of power electronic systems.

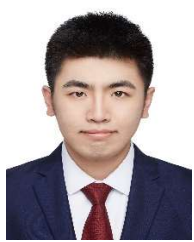
**Jialin Zheng** (Student Member, IEEE) received the B.S. degree in electrical engineering in 2019 from Beijing Jiaotong University, Beijing, China. Since 2019, he has been working toward the Ph.D. degree in electrical engineering at the Department of Electrical Engineering, Tsinghua University, Beijing, China. His research interests include simulation of power electronic systems, modeling of power semiconductor devices, and modeling for high-capacity power electronics devices.

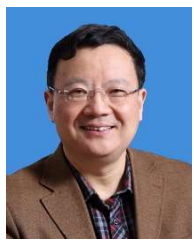
**Zhengming Zhao** (Fellow, IEEE) received the B.S. and M.S. degrees in electrical engineering from Hunan University, Changsha, China, in 1982 and 1985, respectively, and the Ph.D. degree in electrical engineering from Tsinghua University, Beijing, China, in 1991.
He is currently a Professor with the Department of Electrical Engineering, Tsinghua University. His research interests include high-power conversion, power electronics and motor control, and solar energy applications.